\documentclass[25pt,aps,onecolumn,prc,superscriptaddress,noshowpacs,nofootinbib,noshowkeys,floatfix]{article}
\usepackage[dvips]{graphics,graphicx}
\usepackage[colorlinks=true,linktocpage=true,linkcolor=blue,citecolor=blue]{hyperref}
\usepackage[utf8]{inputenc}
\usepackage{bm}
\usepackage{authblk}
\usepackage{amsmath}
\usepackage{graphicx}
\usepackage{amsfonts}
\usepackage{amssymb}
\usepackage{multirow}
\usepackage{hyperref}
\usepackage{lineno}
\usepackage{geometry}
\usepackage[usenames,dvipsnames]{color}
\usepackage{amsmath, amssymb}
\usepackage{multirow}
\usepackage{longtable}
\usepackage{color}
\usepackage[normalem]{ulem}  % \sout{old text} for strikeout

\renewcommand\sout{\bgroup \color{blue} \ULdepth=-.5ex \ULset}
\begin{document}

\title{An insight into strangeness with $\phi$(1020) production in small to large collision systems with ALICE at the LHC}

\author{Sushanta Tripathy (for ALICE collaboration) \thanks{e-mail: Sushanta.Tripathy@cern.ch}}
\affil{Discipline of Physics, School of Basic Sciences, Indian Institute of Technology Indore, Simrol, Indore 453552, India}

%\affil[1]{}
%\date{\today}

\maketitle

\begin{abstract}
Hadronic resonances are unique tools to investigate the interplay of re-scattering and regeneration effects in the hadronic phase of heavy-ion collisions. As the $\phi$ meson has a longer lifetime compared to other resonances, it is expected that its production will not be affected by regeneration and re-scattering processes. Measurements in small collision systems such as proton-proton (pp) collisions provide a necessary baseline for heavy-ion data and help to tune pQCD inspired event generators.  Given that the $\phi$ is a bound state of strange-antistrange quark pair (s$\bar{\rm{s}}$), measurements of its production can contribute to the study of strangeness production. Recent results obtained by using the ALICE detector show that although $\phi$ has zero net strangeness content, it behaves like a particle with open strangeness in small collision systems and the experimental results agree with thermal model predictions in large systems. The production mechanism of $\phi$ is yet to be understood.

We report on measurements with the ALICE detector at the LHC of $\phi$ meson production in pp, p--Pb, Xe--Xe and Pb--Pb collisions. These results are reported for minimum bias event samples and as a function of the charged particle multiplicity or centrality. The results include the transverse momentum ($p_{\rm T}$) distributions of $\phi$ as well as the $\langle p_{\rm T}\rangle$ and particle yield ratios. The $\phi$ effective strangeness will be discussed in relation to descriptions of its production mechanism, such as strangeness canonical suppression, non-equilibrium production of strange quarks and thermal models.

\end{abstract}

\section{Introduction}
\label{int}
Resonances are ideal candidates to probe the hadronic phase formed in heavy-ion collisions due to their short lifetimes. The lifetime of $\phi$ (46.3 fm/{\it c}) is longer compared to that of other hadronic resonances as well as the lifetime of the fireball produced in heavy-ion collisions. Thus it is expected that $\phi$ meson will not be affected by the re-scattering and re-generation processes~\cite{regen} during the hadronic phase. Being the $\phi$ a bound state of a strange-antistrange quark pair (s$\bar{\rm{s}}$), a measurement of its production can help shed light on strangeness production mechanisms. Also, the study of $\phi$ in small colliding systems helps in the search for the onset of collectivity and provides a necessary baseline for heavy-ion collisions. 

This article focuses on measurements of $\phi$ production with the ALICE detector at the LHC in pp collisions at $\sqrt{s}$ = 0.9, 2.76, 5.02, 7, 8 and 13 TeV, p--Pb collisions at 5.02 and 8.16 TeV, Xe--Xe collisions at $\sqrt{s_{\rm{NN}}}$ = 5.44 TeV and Pb--Pb collisions at $\sqrt{s_{\rm{NN}}}$~=~2.76~and~5.02~TeV. In particular, $p_\mathrm{T}$ spectra at different energies and colliding systems as well as $p_\mathrm{T}$-integrated particle ratios to long-lived hadrons are compared for minimum bias collisions and as a function of the charged particle multiplicity ($\langle \mathrm{d}N_{\mathrm{ch}}/\mathrm{d}\eta\rangle$). In this paper, we aim at addressing one of the major questions, namely whether $\phi$ behaves like a non-strange or strange particle. The strangeness of $\phi$ will be discussed in relation to its production mechanism, such as strangeness canonical suppression, non-equilibrium production of strange quarks and thermal models.

%% The Appendices part is started with the command \appendix;
%% appendix sections are then done as normal sections
%% \appendix

\section{$\phi$ meson reconstruction and $p_\mathrm{T}$ spectra}
\label{rec}

\begin{figure}[ht!]
\centering
\includegraphics[height=13em]{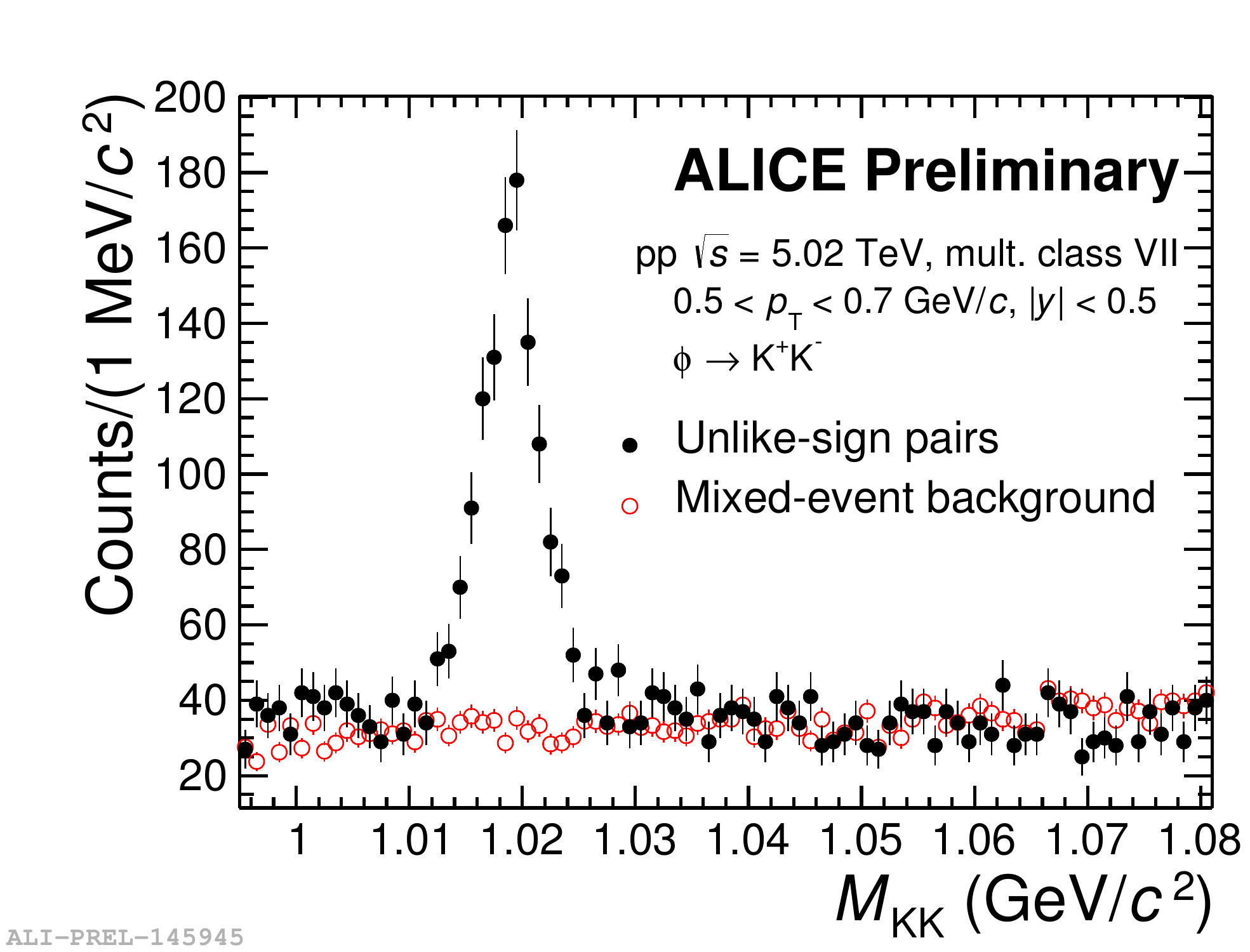}
\includegraphics[height=13em]{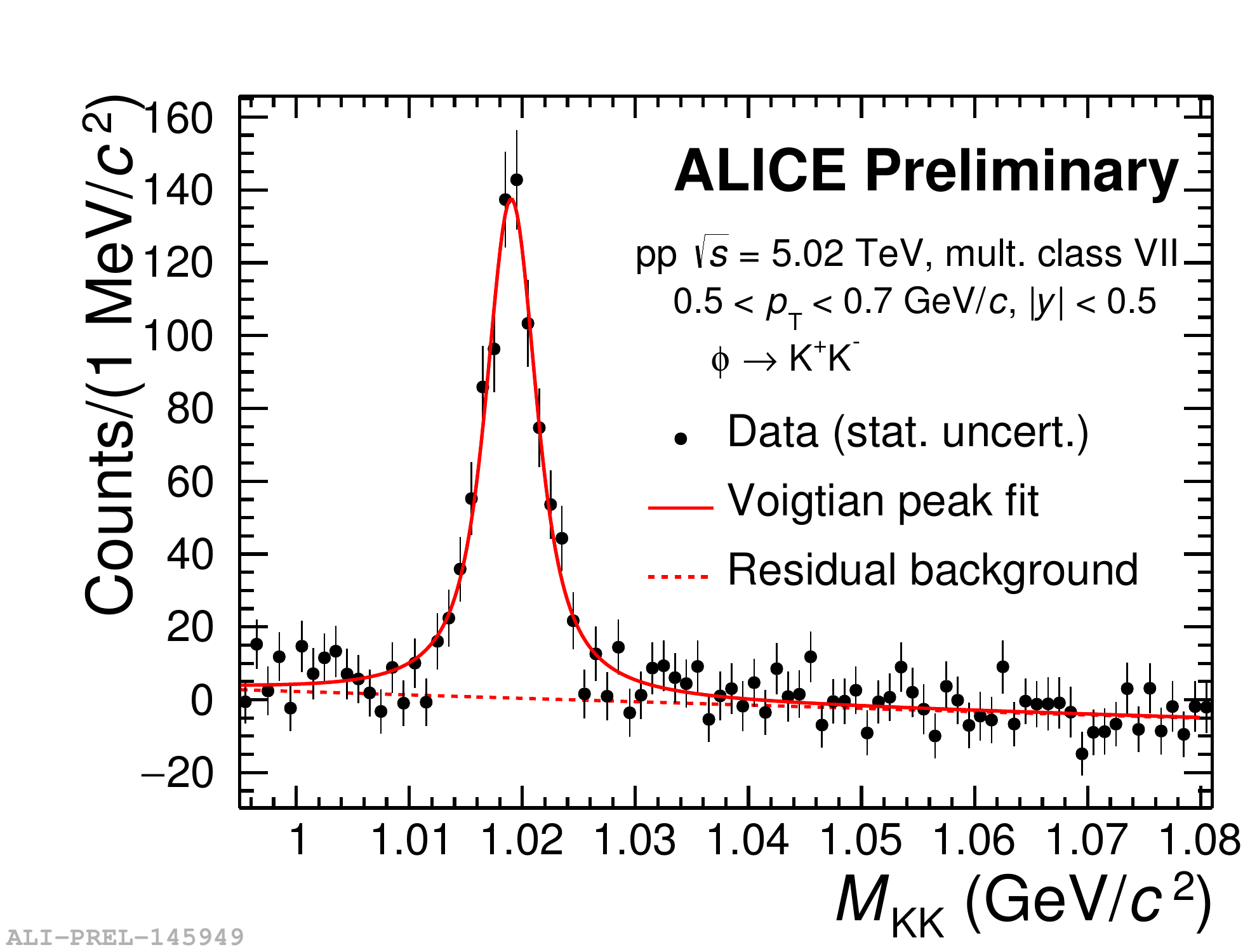}
\caption[]{Invariant-mass distribution for the $\phi$ in pp collisions at $\sqrt{s}$~=~5.02 TeV in one of the measured $p_{\mathrm{T}}$ ranges in V0M Multiplicity class VII. Left: the unlike-charge invariant-mass distribution with mixed-event backgrounds. Right: Invariant-mass distribution after subtraction of the mixed-event background with a Voigtian fit to describe the peak of the $\phi$ and the residual background.}
\label{figa}
\end{figure}

The $\phi(1020)$ is reconstructed at mid-rapidity ($|y|<$ 0.5) through an invariant mass analysis via its hadronic decay channel~\cite{2017,2012} into K$^{+}$K$^{-}$ (branching ratio: 49.2\%)~\cite{pdg}. Figure~\ref{figa} shows the invariant-mass distribution for the $\phi$ in pp collisions at $\sqrt{s}$ = 5.02 TeV in the $p_{\mathrm{T}}$ range 0.5 $<$ $p_{\mathrm{T}}$ $<$ 0.7 GeV/$c$ in V0M Multiplicity class VII. The left plot of Fig.~\ref{figa} shows the unlike-charge invariant-mass distribution with a combinatorial background.
The event mixing and like-sign techniques are used to estimate the combinatorial background and after combinatorial background subtraction a residual background remains as shown in the right plot of Fig. 1, together with a fit used to describe the peak of $\phi$ and the residual background. The latter is mainly due to mis-identified particle decay products or from other sources of correlated pairs (e.g. mini-jets). The $\phi(1020)$ peak is fitted with a Voigtian function, a convolution of Breit-Wigner and Gaussian functions~\cite{2017,2012}. For some cases, the $\phi$ peak is fitted without any combinatorial background subtraction when the combinatorial background shows large statistical fluctuation.

\begin{figure}[ht!]
\centering
\includegraphics[height=19.3em]{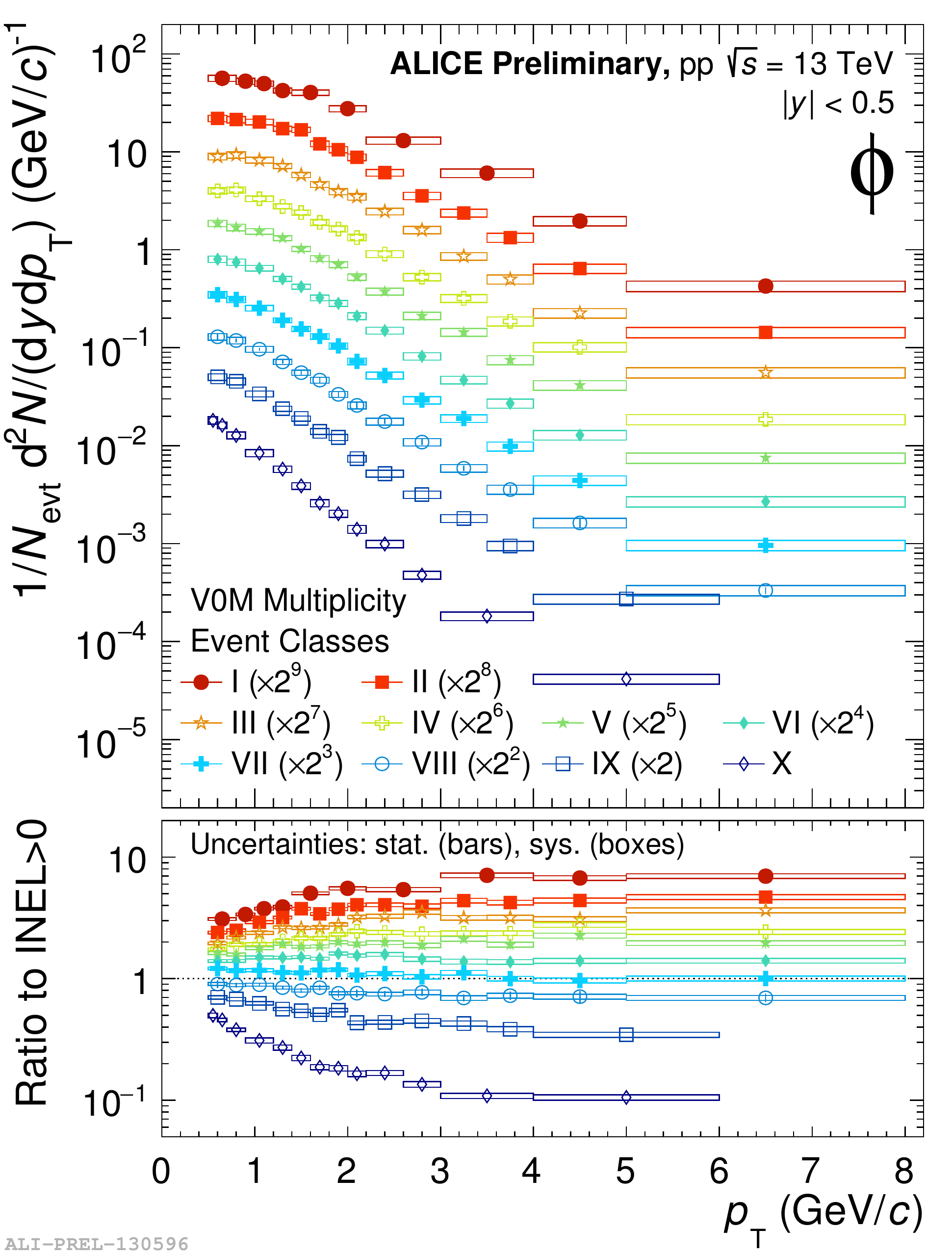}
\includegraphics[height=19.3em]{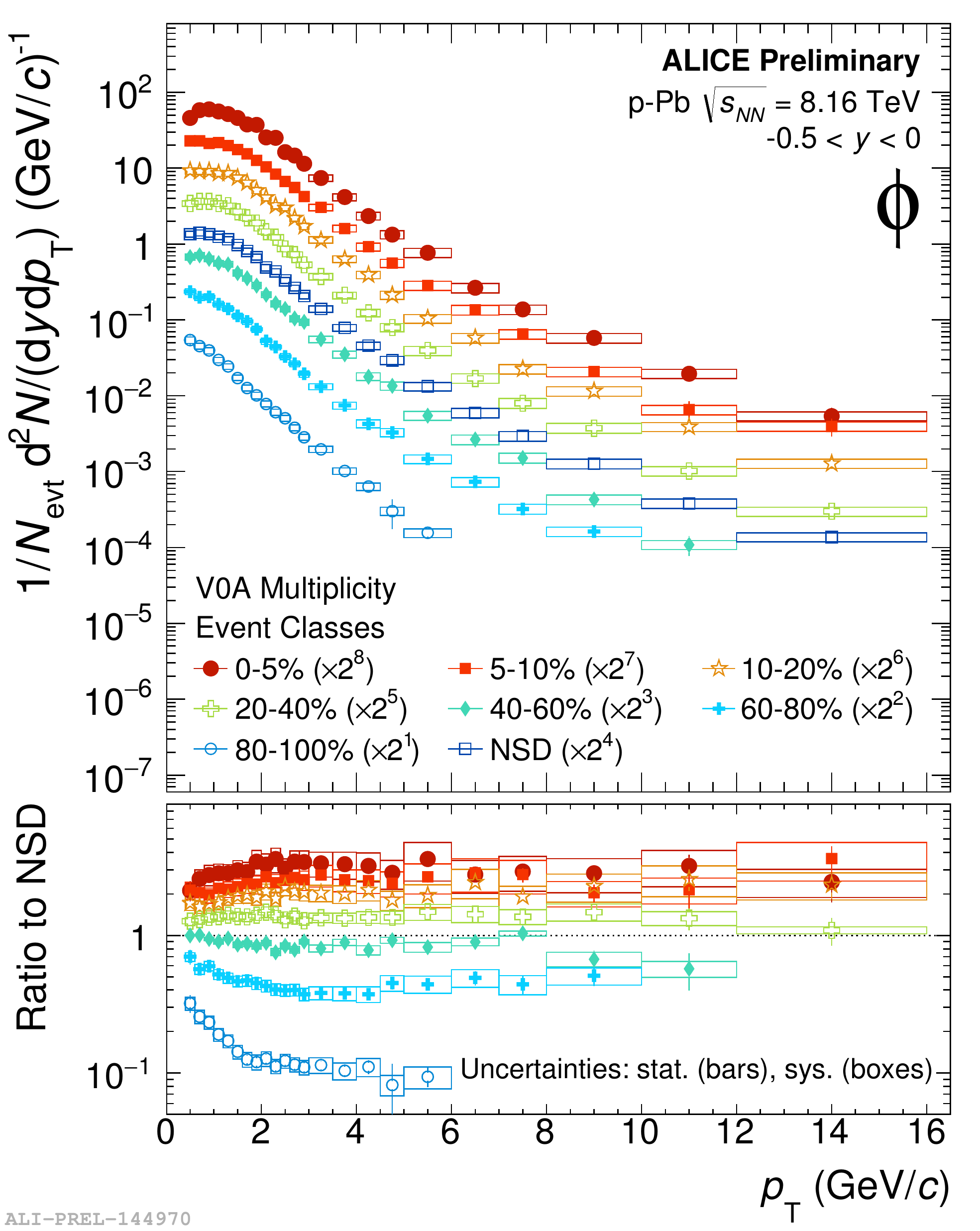}
\caption[]{$p_{\mathrm{T}}$  spectra of $\phi$ mesons in pp collisions at $\sqrt{s}$ = 13 TeV (left) and p-Pb collisions at $\sqrt{s_{\rm{NN}}}$ = 8.16 TeV (right) in different multiplicity classes. In the bottom panels of the figure the ratios of the $p_{\mathrm{T}}$ spectra to the multiplicity-integrated $p_{\mathrm{T}}$ spectrum is reported.}
\label{fig2}
\end{figure}

\begin{figure}[ht!]
\centering
\includegraphics[height=16.2em]{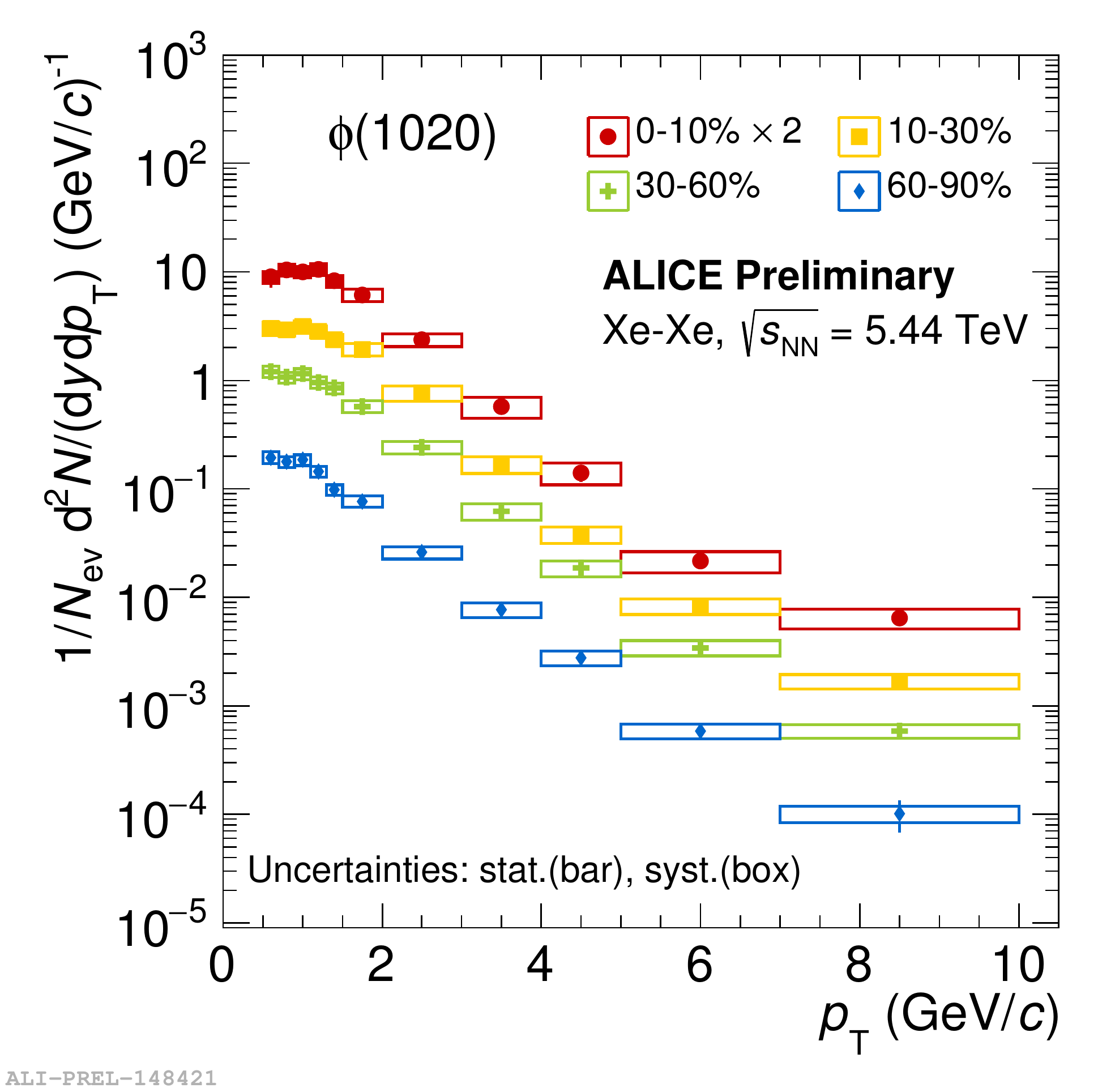}
\includegraphics[height=16.1em]{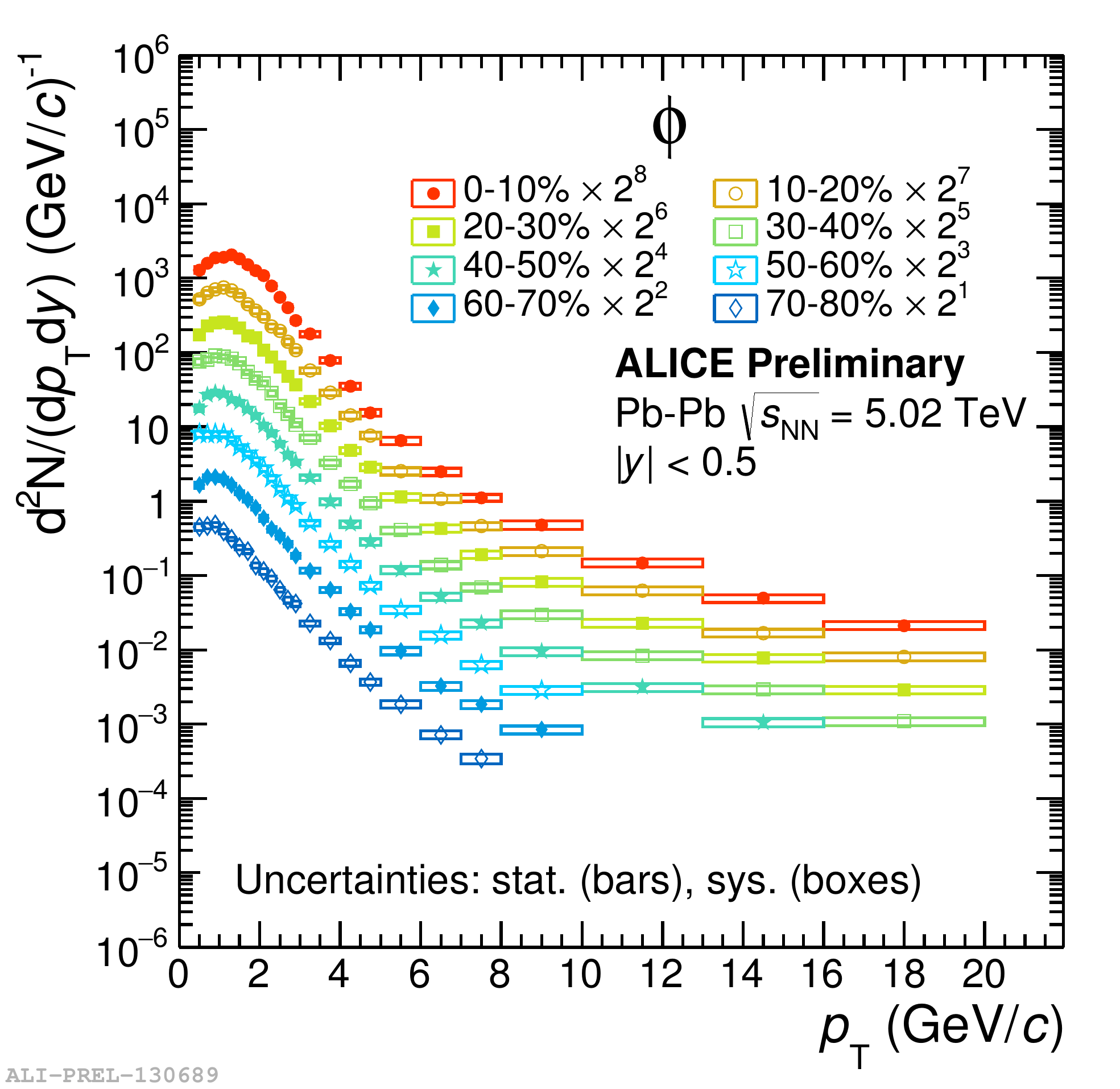}
\caption[]{$p_{\mathrm{T}}$ spectra of $\phi$ mesons in Xe-Xe collisions at $\sqrt{s_{\rm{NN}}}$ = 5.44 TeV (left) and in Pb-Pb collisions at $\sqrt{s_{\rm{NN}}}$ = 5.02 TeV (right) in different centrality classes.}
\label{fig3}
\end{figure}

 In each $p_{\mathrm{T}}$ intervals, raw yields are obtained from the fit to the signal peak and then corrected for the detector efficiency $\times$ acceptance and the branching ratio to determine the final $p_{\mathrm{T}}$ spectrum. Figure~\ref{fig2} shows the $p_{\mathrm{T}}$  spectra of $\phi$ mesons in pp collisions at $\sqrt{s}$ = 13 TeV (left) and p-Pb collisions at $\sqrt{s_{\rm{NN}}}$ = 8.16 TeV (right) in different V0 multiplicity classes. The lower panels show the ratio of the $p_{\mathrm{T}}$ spectra to the 0-100\% $p_{\mathrm{T}}$ spectrum. Evolution of $p_{\mathrm{T}}$-spectra is observed at low-$p_{\mathrm{T}}$. For high-$p_{\mathrm{T}}$, the slopes of the spectra in different multiplicity classes seem to be similar to those observed in minimum bias pp collisions. Figure~\ref{fig3} shows the $p_{\mathrm{T}}$ spectra of $\phi$ mesons in Xe-Xe collisions at $\sqrt{s_{\rm{NN}}}$ = 5.44 TeV (left) and Pb-Pb collisions at $\sqrt{s_{\rm{NN}}}$ = 5.02 TeV (right) for different centrality classes. 

\section{Results and Discussion}
\label{rec}

\begin{figure}[ht!]
\centering
\includegraphics[height=18.9em]{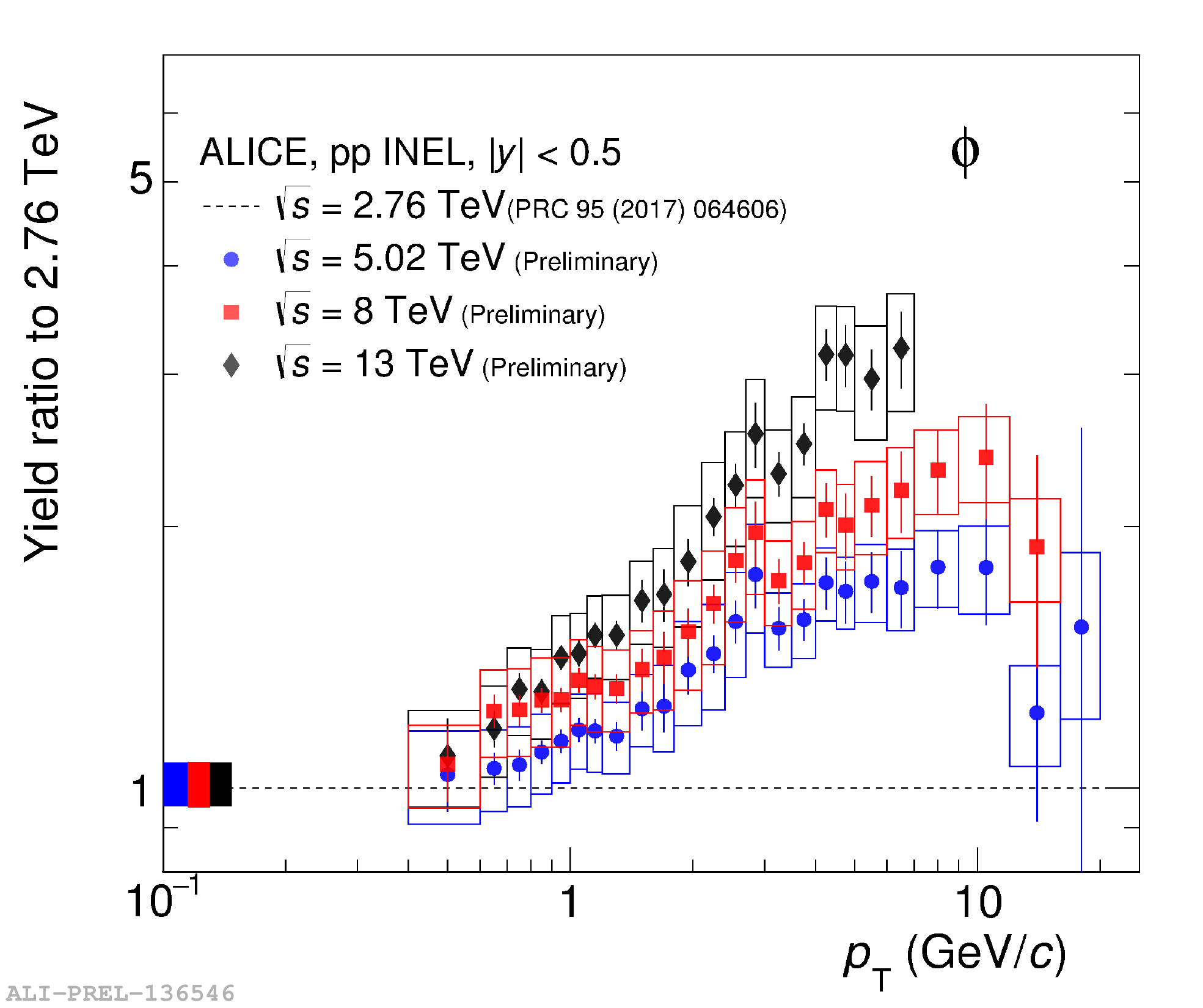}
\caption[]{Ratios of $p_{\mathrm{T}}$ spectra of $\phi$ in inelastic pp collisions at various center of mass energies to the spectrum obtained in pp collisions at $\sqrt{s}$ = 2.76 TeV. Statistical uncertainties are represented by bars and systematic uncertainties are represented by boxes.}
\label{fig4}
\end{figure}

Figure~\ref{fig4} shows ratios of $p_{\mathrm{T}}$ spectra of $\phi$ in inelastic pp collisions at various center-of-mass energies to the spectrum obtained in pp collisions at $\sqrt{s}$ = 2.76 TeV. These ratios indicate that from 1-2 GeV/$\it c$, the yields increase as a function of collision energy, but the production at low $p_{\mathrm{T}}$ does not strongly depend on collision energy.

\begin{figure}[ht!]
\centering
\includegraphics[height=15.3em]{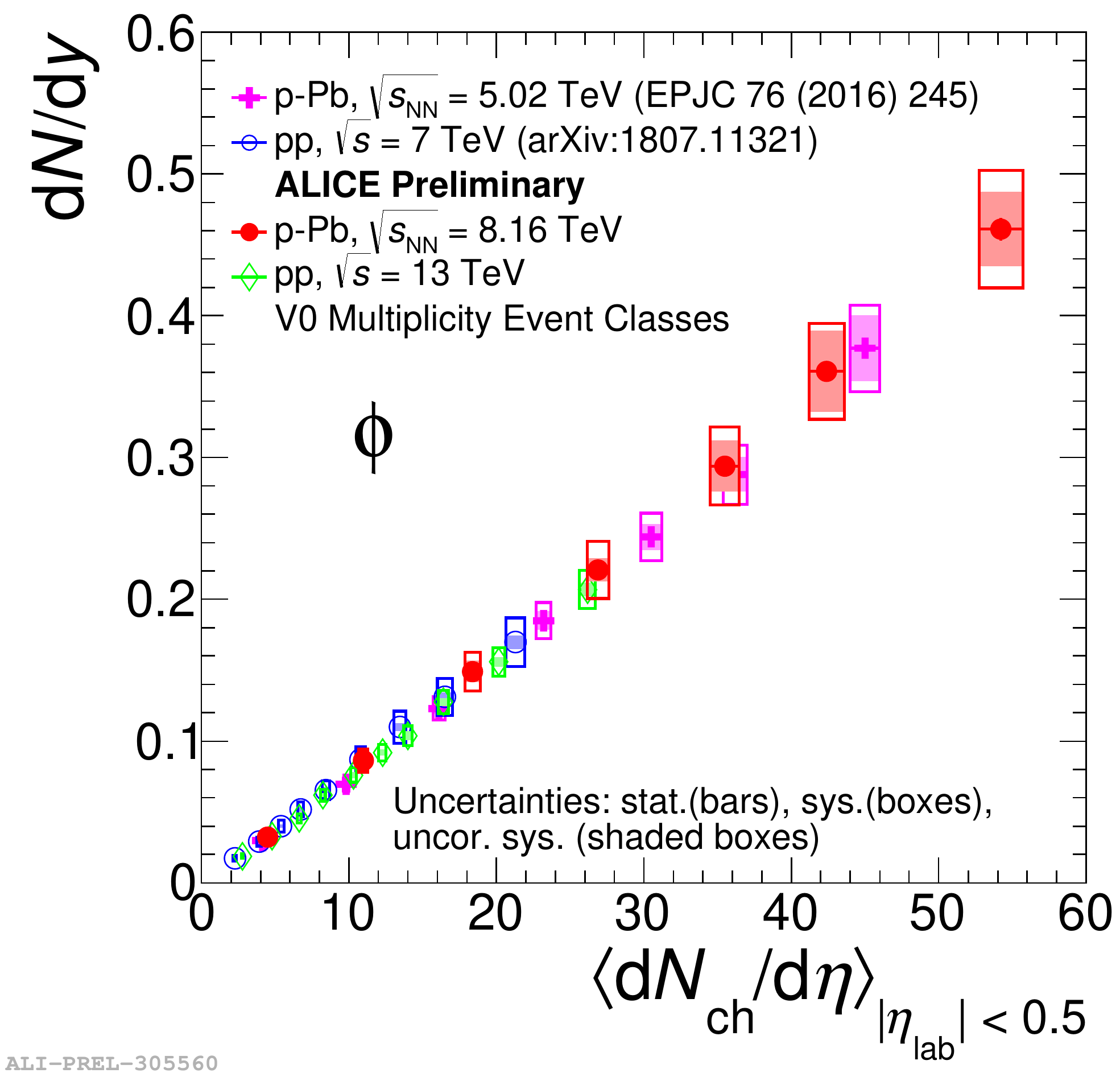}
\includegraphics[height=15.3em]{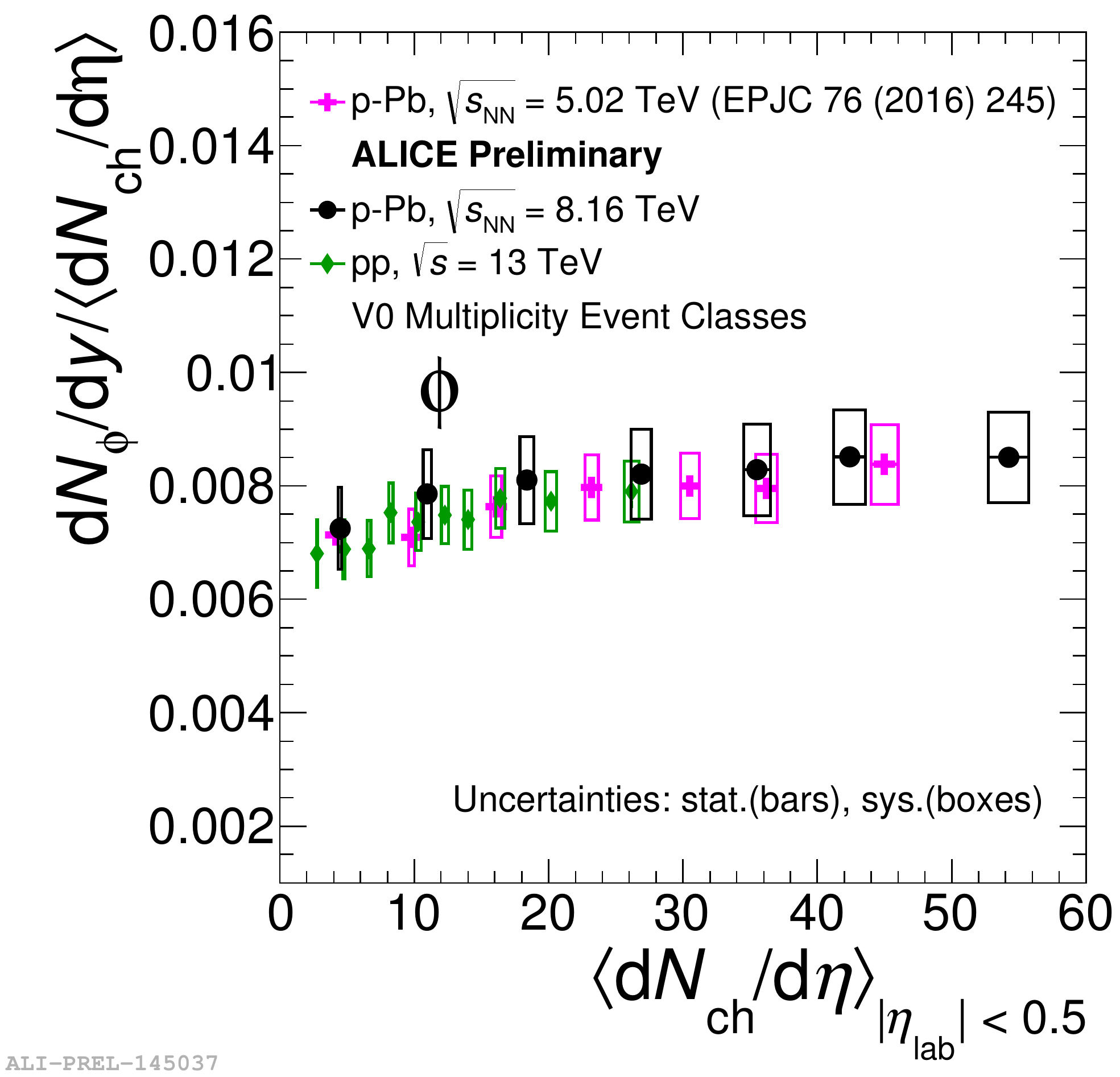}
\caption[]{Left: $\mathrm{d}N/\mathrm{d}y$ of $\phi$ as a function of charged particle multiplicity in pp collisions at $\sqrt{s}$ = 7 and 13 TeV and p--Pb collisions at $\sqrt{s_{\rm{NN}}}$ = 5.02 and 8.16 TeV. Right: ($\mathrm{d}N/\mathrm{d}y$)/$\langle \mathrm{d}N_{\mathrm{ch}}/\mathrm{d}\eta\rangle$ for $\phi$ as a function of average charged particle multiplicity in pp collisions at $\sqrt{s}$ = 13 TeV and p--Pb collisions at $\sqrt{s_{\rm{NN}}}$ = 5.02 and 8.16 TeV. Statistical uncertainties are represented by bars and systematic uncertainties are represented by boxes.}
\label{fig5}
\end{figure}

The left panel of Fig.~\ref{fig5} shows the integrated yield of $\phi$ in pp collisions at $\sqrt{s}$~=~7~and~13~TeV and p--Pb collisions at $\sqrt{s_{\rm{NN}}}$ =~5.02 and 8.16 TeV. The integrated yield shows a linear increase as a function of charged-particle multiplicity for both pp and p-Pb collisions. The right panel of Fig.~\ref{fig5} shows the $\phi$ yield normalized by the $\langle \mathrm{d}N_{\mathrm{ch}}/\mathrm{d}\eta\rangle$ value as a function of average charged particle multiplicity in pp collisions at $\sqrt{s}$ = 13 TeV and in p--Pb collisions at $\sqrt{s_{\rm{NN}}}$ = 5.02 and 8.16 TeV. The ratio is independent of collision energy, which suggests that the event multiplicity drives the particle production, irrespective of collision system type and energy. 

\begin{figure}[ht!]
\centering
\includegraphics[height=14.2em]{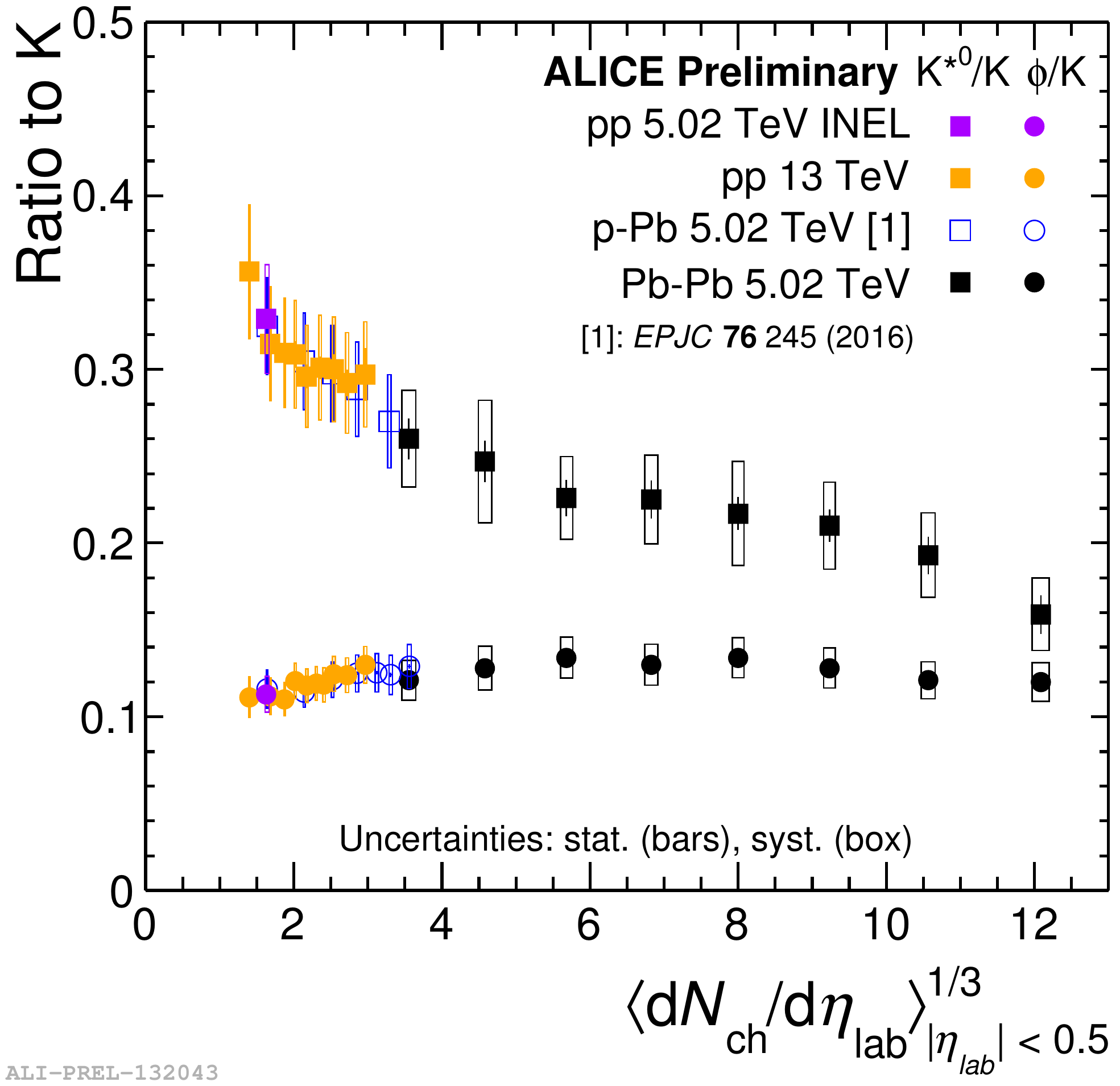}
\includegraphics[height=14.6em]{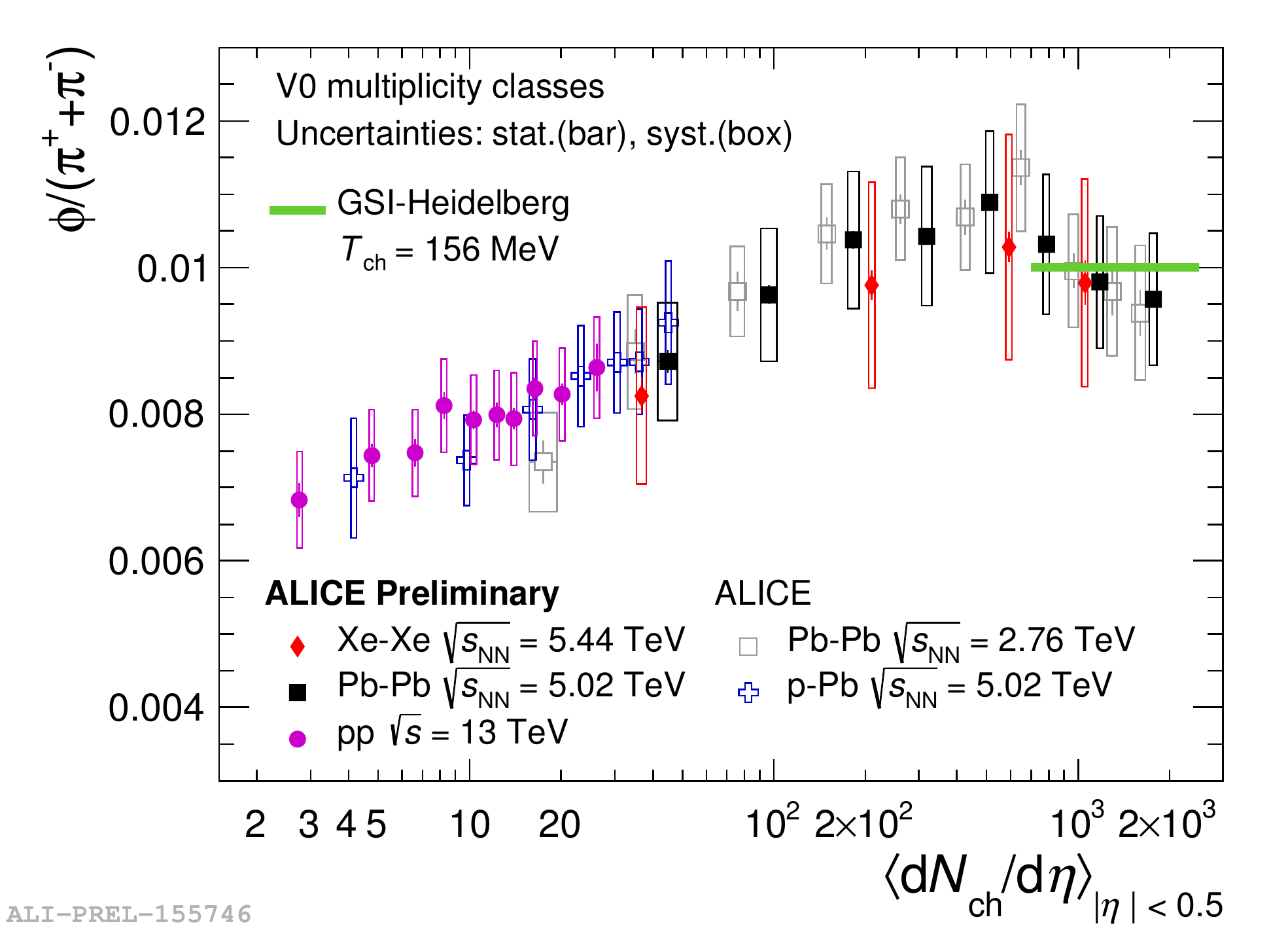}
\includegraphics[height=14.2em]{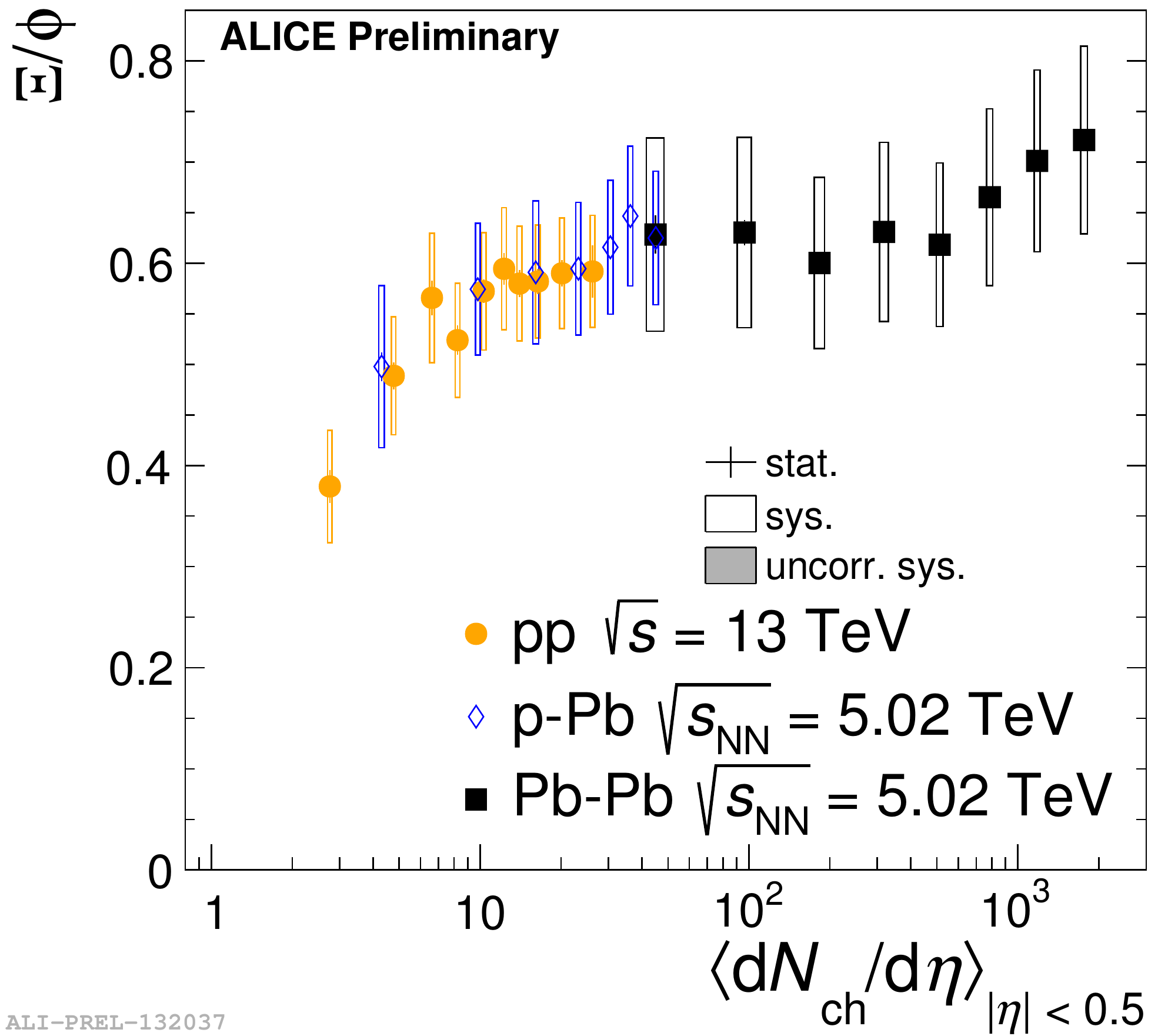}
\caption[]{Ratios of the $p_{\mathrm{T}}$-integrated yield of $\phi$ and K$^{*0}$ relative to K (top left), $\phi$ relative to $\pi$ (top right) and $\Xi$ relative to $\phi$ (bottom) as a function of the charged particle multiplicity in different collision systems for different center-of-mass energies.}
\label{fig6}
\end{figure}

The top left panel of Fig.~\ref{fig6} shows the yield ratios of $\phi$ and K${^{*0}}$ to kaons as a function of charged particle multiplicity for different colliding systems at different collision energies. As the lifetime of K$^{*0}$ is almost 10 times shorter compared to $\phi$, it is expected that K${^{*0}}$ is affected by the re-generation and/or re-scattering processes in a long-lasting hadronic phase of the expanding system. A decreasing trend in the K$^{*0}$/K ratio is observed, suggesting that the re-scattering mechanism dominates over regeneration. As expected, the $\phi$/K ratio remains fairly flat, which indicates that either the regeneration and re-scattering are balanced or $\phi$ decays after the hadronic phase without being affected by these processes. The top right panel of Fig.~\ref{fig6} shows the $\phi$/$\pi$ ratio as a function of $\langle \mathrm{d}N_{\mathrm{ch}}/\mathrm{d}\eta\rangle$. The production of $\phi$ in Pb--Pb and Xe--Xe collisions is well described by a grand-canonical thermal model (GSI-Heidelberg)~\cite{Stachel:2013zma}, while for small systems (pp and p--Pb collisions) the increase of the $\phi/\pi$ ratio with multiplicity is in contrast to the expectation from strangeness canonical suppression~\cite{Acharya:2018orn}. This behavior favors the non-equilibrium production of $\phi$ and/or strange particles. The bottom panel of Figure~\ref{fig6} shows the $\Xi$/$\phi$ ratio as a function of $\langle \mathrm{d}N_{\mathrm{ch}}/\mathrm{d}\eta\rangle$. The $\Xi/\phi$ ratio remains fairly flat or slightly increases across a wide multiplicity range. In addition, a multiplicity dependence of the ratio is observed, particularly at low multiplicities. Comparing the $\phi$ with particles with strange quark content 1 or 2, we observe that the $\phi$ behaves like a particle with open strangeness~\cite{Tripathy:2018ehz}.

\section{Summary}
\label{sum}
ALICE has studied $\phi$ production as a function of collision energy and charged particle multiplicity in different colliding systems. The event multiplicity seems to drive the production of hadrons, including $\phi$ production, irrespective of collision energy for pp and p-Pb collisions at the LHC. The $\phi$/K ratio remains rather flat across a wide range of multiplicity and across colliding systems, which indicates that either regeneration and re-scattering are balanced or that the $\phi$ decays after the hadronic phase in Pb--Pb collisions and is not affected by re-scattering and regeneration. The latter seems to be the likely scenario as $\phi$ has an almost 10~times longer lifetime than K$^{*0}$. Looking at the $\phi$/$\pi$, $\phi$/K and $\Xi/\phi$ ratios, the $\phi$ meson seems to show a similar behavior to that of particles with open strangeness.

\section{Acknowledgements}
ST acknowledges the financial support by DST-INSPIRE program of the Government of India.

%% References
%%
%% Following citation commands can be used in the body text:
%% Usage of \cite is as follows:
%%   \cite{key}         ==>>  [#]
%%   \cite[chap. 2]{key} ==>> [#, chap. 2]
%%

%% References with BibTeX database:

%\bibliographystyle{elsarticle-num}
%\bibliography{<your-bib-database>}

%% Authors are advised to use a BibTeX database file for their reference list.
%% The provided style file elsarticle-num.bst formats references in the required Procedia style

%% For references without a BibTeX database:

\end{document}